\documentclass[]{spie} 
\usepackage{amsmath,amsfonts,amssymb}
\usepackage{graphicx}
\usepackage{subcaption}
\usepackage[colorlinks=true, allcolors=blue]{hyperref}
\usepackage{multirow}
\usepackage[perpage]{footmisc}
\usepackage{enumitem}

\title{First results for second generation SiSeRO CCD devices}

\author[a]{Abigail Y. Pan}
\author[a]{Declan O'Neill}
\author[b]{Kevan Donlon}
\author[a]{Peter Orel}
\author[a]{Sven Herrmann}
\author[a]{Steven Allen}
\author[c]{Marshall W. Bautz}
\author[a]{Tanmoy Chattopadhyay}
\author[b]{Michael Cooper}
\author[c]{Catherine E. Grant}
\author[c]{Jill Juneau}
\author[b]{Chris Leitz}
\author[c]{Beverly LaMarr}
\author[c]{Eric D. Miller}
\author[a]{Glenn Morris}
\author[a]{Tonya L. Peshel}
\author[a]{Artem Poliszczuk}
\author[c]{Gregory Prigozhin}
\author[b]{Ilya Prigozhin}
\author[a]{Haley R. Stueber}
\author[b]{Keith Warner}
\affil[a]{Kavli Institute for Particle Astrophysics and Cosmology, Stanford University, 452 Lomita Mall, Stanford, CA 94305, USA}
\affil[b]{MIT Lincoln Laboratory, Lexington, MA, USA}
\affil[c]{MIT Kavli Institute for Astrophysics and Space Research, Massachusetts Institute of Technology, Cambridge, MA, USA}

\authorinfo{Abigail Y. Pan: E-mail: apan5@stanford.edu}

\begin{document} 
\maketitle

\begin{abstract}
The Astro2020 Decadal recommended the development of a suite of next generation astronomical observatories spanning the X-ray to near-IR spectrum. These programs require fast, extremely low noise detectors to fulfill their science goals. To address this technology gap, Stanford X-ray Astronomy and Observational Cosmology (XOC) group, MIT Lincoln Laboratory (MIT-LL), and MIT Kavli Institute (MKI) are advancing Single electron Sensitive Read Out (SiSeRO), a multiband detector technology capable of achieving substantially sub-electron noise via Repetitive Non-Destructive Readout (RNDR). We present initial results for our second generation SiSeRO CCDs. We also discuss our test bed, including a readout electronics system capable of accommodating all second-generation SiSeRO CCD variants utilizing the XOC-designed Multi-Channel Readout Chip (MCRC) ASIC.
\end{abstract}

%%Manuscript format, template, SPIE Proceedings, LaTeX}

\section{Introduction}
\label{sec:intro}  % \label{} allows reference to this section'

One of the top priorities recommended by the Astro2020 Decadal Survey report \textit{(Pathways to Discovery in Astronomy and Astrophysics for the 2020s)} is the development of a suite of complementary next-generation flagship observatories across multiple wavelength bands. The report highlighted ground-based observatories in the visible to infrared, such as the Extremely Large Telescopes \cite{johnsGiantMagellanTelescope2012, sandersThirtyMeterTelescope2013, neichel2018overvieweuropeanextremelylarge}, as well as space missions in the UV/visible/near-IR, such as Habitable Worlds Observatory \cite{feinbergHabitableWorldsObservatory2024}, and X-ray regime, such as a Lynx-like X-ray flagship \cite{gaskinLynxXRayObservatory2019}. These observatories will deploy spectro-imaging instruments that require large-format, small-pixel detectors to achieve high spatial resolution, fast readout speeds to reduce the effect of photon pileup\cite{davisEventPileupChargecoupled2001} and background events, and extremely low noise to achieve high sensitivity for faint sources and soft energies. 

\begin{figure} [ht]
\centering
\begin{subfigure}{0.55\textwidth}
    \includegraphics[width=\textwidth]{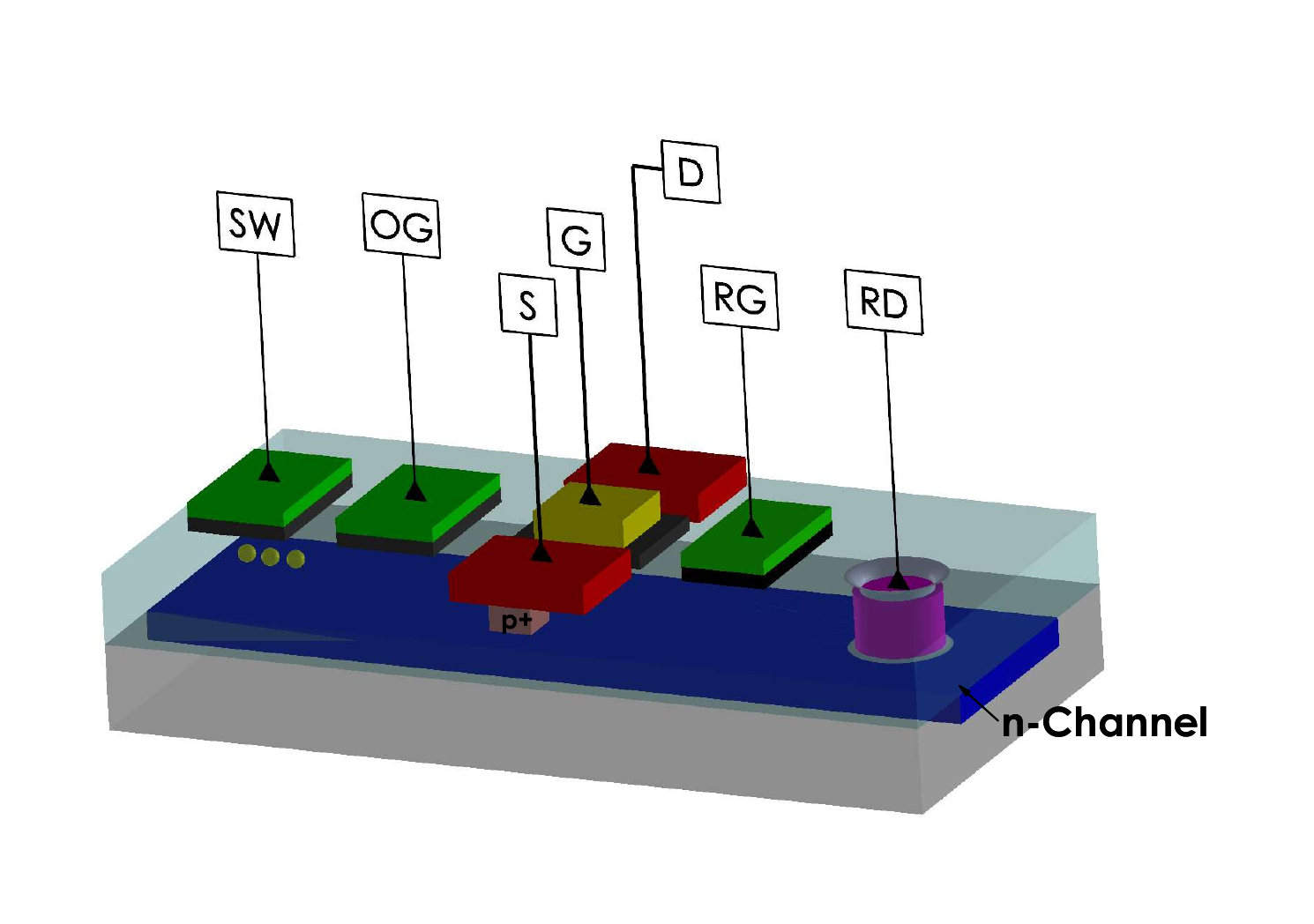}
\end{subfigure}
\begin{subfigure}{0.35\textwidth}
    \includegraphics[width=\textwidth]{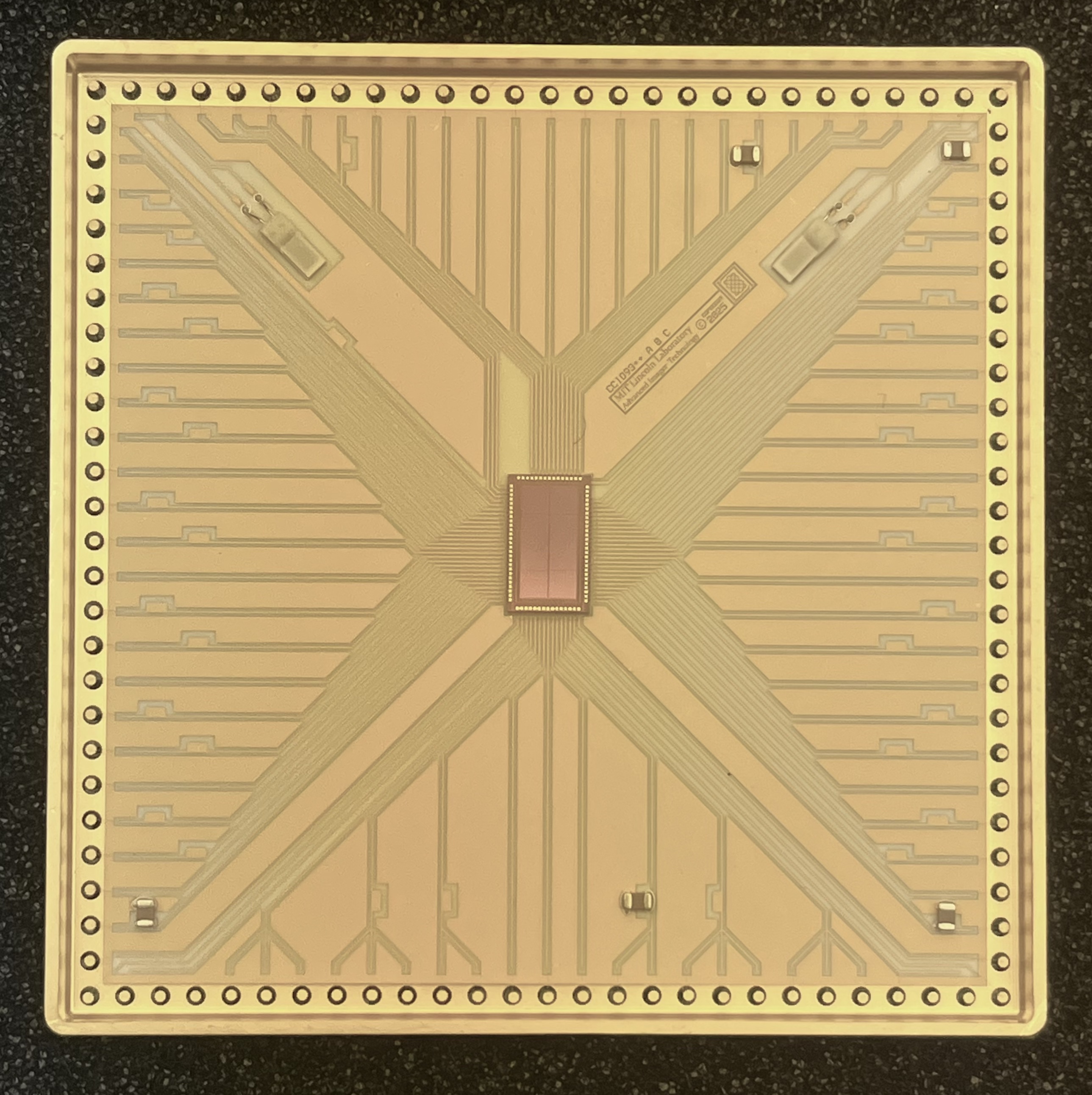}
\end{subfigure}
\caption{Left: Schematic of the SiSeRO output stage.\cite{chattopadhyayDevelopmentCharacterizationFast2022}. The p-MOSFET transistor (S: source, G: gate, D: drain terminals) straddles the output transfer channel. A internal gate is located beneath the transistor. When a charge packet is transferred from the summing well (SW) and output gate (OG) into the internal gate, the transistor drain current is modulated. The internal channel is emptied via the reset gate (RG) and reset drain (RD). Right: A second generation CCID93++B parallel SiSeRO device.}
\label{fig:sisero_diagram}  

\end{figure}

The Single-electron Sensitive Read Out (SiSeRO) is a novel detector output stage that is targeting this technology gap \cite{chattopadhyayFirstResultsSiSeRO2022}. These devices are being advanced by a collaboration between the X-ray Astronomy and Observational Cosmology (XOC) group at Stanford, MIT Lincoln Laboratory (MIT-LL) and the MIT Kavli Institute (MKI). The SiSeRO replaces the floating diffusion gate found in traditional charge coupled device (CCD) output stages with an internal gate underneath a p-type metal oxide semiconductor field effect transistor (p-MOSFET). This design improves gain and noise by reducing the parasitic capacitance of the output stage. Figure \ref{fig:sisero_diagram} shows a simplified schematic of the output design. Charge is measured in drain current mode\cite{wolfelSubelectronNoiseMeasurements2006}: the current in the p-MOSFET transistor is modulated when a charge packet is transferred to the internal gate; this current is then read out through the source or drain terminal. Because this readout process preserves the original charge packet, multiple independent measurements of the charge can be made in a process known as repetitive non-destructive readout (RNDR). These measurements can reduce the read noise by a factor of $\sqrt{\mathrm{N}}$ (where N is the number of RNDR cycles), enabling SiSeROs to achieve sub-electron noise.

The first generation of SiSeRO detectors, dubbed the CCID93, have demonstrated sub-electron capabilities at competitive readout speeds: at an RNDR cycle rate of 1.6 $\mu$s (625 kHz), we were able to achieve 1 $e^-_{\mathrm{RMS}}$ noise in 15 cycles and 0.49 $e^-_{\mathrm{RMS}}$ noise in 57 cycles, a rate of 42 kpix/sec and 10 kpix/sec respectively\cite{chattopadhyayDemonstratingSubelectronNoise2024}. In addition, these detectors have demonstrated excellent noise performance at larger numbers of RNDR cycles, showing no evidence of non-Poisson behavior out the maximum 200 RNDR cycles tested \cite{peshel2026}. We measure a gain of 800 pA/$e^-$ and a spectral resolution of $\sim 132$ eV (Full Width at Half Maximum (FWHM)) for a 5.9 keV Mn K-$\alpha$ emission line. \cite{chattopadhyayDemonstratingRepetitiveNondestructive2024}.

In this manuscript, we will discuss the design and development of the second generation of SiSeRO devices, their dedicated test bed, and the current progress of the commissioning process. 

\section{Second-generation SiSeRO Devices}
\label{sec:93pp}

To advance the SiSeRO output technology, we have developed a family of second-generation devices, dubbed the CCID93++, to explore different multiple-output SiSeRO readout configurations. Figure \ref{fig:sisero_diagram} shows a fully packaged second-generation CCID93++ detector. 

\subsection{Parallel SiSeRO Readout: The CCID93++B}
\begin{figure} [t]
\centering
\begin{subfigure}{0.45\textwidth}
    \raisebox{10mm}{\includegraphics[width=\textwidth]{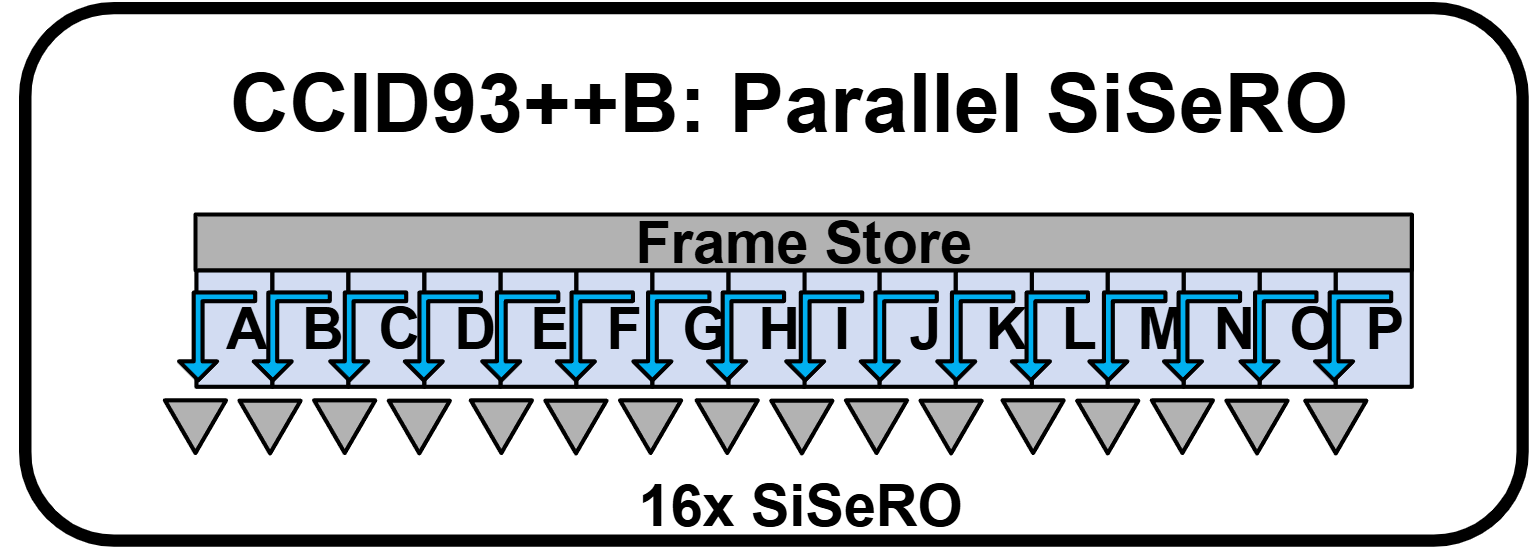}}
\end{subfigure}
\begin{subfigure}{0.40\textwidth}
    \includegraphics[width=\textwidth]{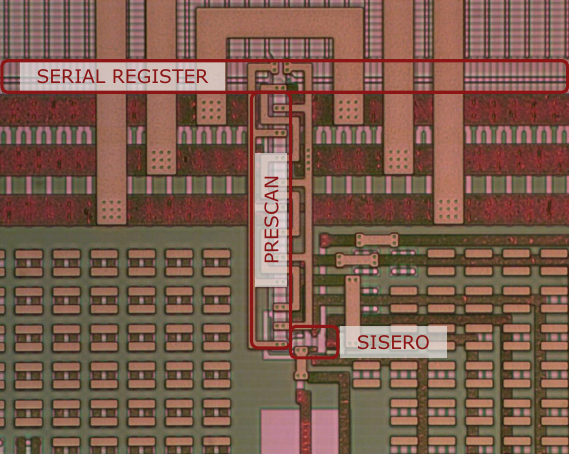}
    % \label{fig:sisero_diagram} 
\end{subfigure}
\caption{Left: The CCID93++B readout schematic\cite{donlonDirectionsAdvancementMITLL2024}. Sixteen SiSeRO outputs are operated in parallel to provide a 16$\times$ increase in readout speed. The outputs are divided into four groups of four output variants for parallelized output development. Right: A photograph of a CCID93++B SiSeRO output. Each output reads out 5 pre-scan pixels and 32 serial pixels.}
\label{fig:CCID93pp_B}  

\end{figure}

The first of these devices is the CCID93++B, which operates 16 SiSeRO outputs in parallel. Figure \ref{fig:CCID93pp_B} shows a schematic of the readout configuration, as well as a photograph of one of the detector outputs. This device shares the same 512$\times$512, 8$\mu$m pixel imaging area as the first-generation CCID93 detectors. The 512 serial pixels are divided across 16 SiSeRO outputs, placed at a 256 $\mu$m pitch, which provides a 16$\times$ increase in readout speed and similar noise compared to a single-channel detector. Each detector can also be operated in RNDR mode to reduce the read noise for an increase in readout time. 

The 16 SiSeROs are divided into four groups to test four different output designs\cite{donlonDirectionsAdvancementMITLL2024}: 
\begin{enumerate}[noitemsep]
    \item The legacy first-generation SiSeRO output.
    \item Shifted internal gate location.
    \item Increased separation between the p-MOSFET source/drain terminals and surrounding channel stop.
    \item Reduced p-MOSFET gate resistance. 
\end{enumerate}

These variants are intended to increase the output gain and speed and lower the intrinsic noise of the SiSeRO output stage. Further details about the variant design can be found in Donlon et al. 2025 \cite{donlonDirectionsAdvancementMITLL2024}. %In addition, we are exploring the effect of different dopant dosages in the CCID93++ wafer fabrication process. 
The design of the CCID93++B allows for a more systematic and efficient way to test many SiSeRO variants simultaneously.

\begin{figure} [b]
\centering
\begin{subfigure}{0.45\textwidth}
    \raisebox{10mm}{\includegraphics[width=\textwidth]{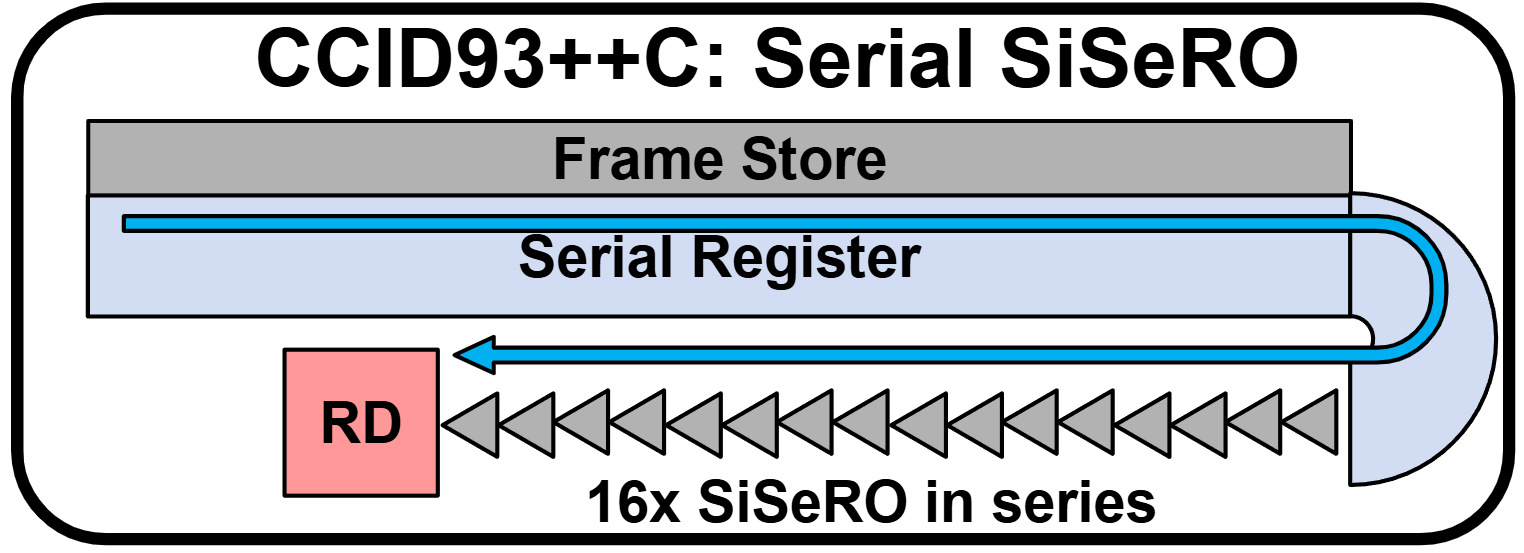}}
\end{subfigure}
\begin{subfigure}{0.40\textwidth}
    \includegraphics[width=\textwidth]{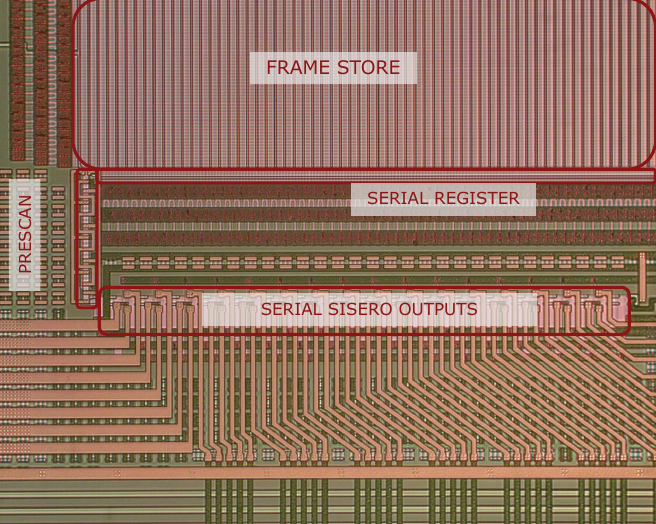}
    % \label{fig:sisero_diagram} 
\end{subfigure}
\caption{Left: The CCID93++C readout schematic\cite{donlonDirectionsAdvancementMITLL2024}. Sixteen identical SiSeROs in series provide a 4$\times$ reduction in read noise. During readout, each charge packet is passed non-destructively along the 16 outputs. This is the first SiSeRO device operated in a multi-amplifier sensor (MAS) configuration. Right: A photograph of the CCID93++C output. }
\label{fig:CCID93pp_C}  

\end{figure}

\subsection{Serial SiSeRO Readout: The CCID93++C}

The second device is the CCID93++C, which operates 16 SiSeRO outputs in series. This is analogous to a multi-amplifier sensor (MAS) CCD\cite{bottiSingleQuantumMeasurementMultipleAmplifier2024}. Figure \ref{fig:CCID93pp_C} shows a schematic of the readout configuration and a photograph of the detector output. This design shares the same form factor of the imaging area and frame store as the CCID93++B. Leveraging the RNDR capabilities of the SiSeRO, each charge packet is passed non-destructively through all sixteen outputs during readout. Subsequent pixels are through the same chain, so that 16 pixels will be measured at a given time. This results in a $\sqrt{16} = 4\times$ reduction in read noise compared to a single-output device operating at the same frame rate. The outputs have a 27$\mu$m pitch, with dedicated transfer gates in between to enhance charge transfer and prevent charge spilling between pixels. For an additional time penalty, traditional RNDR charge transfer can be used to decrease the read noise even further, again with an associated reduction in readout speed.

\subsection{SiSeRO Active Pixel Sensor Matrix}

\begin{figure} [h]
\centering
\begin{subfigure}{0.54\textwidth}    {\includegraphics[width=\textwidth]{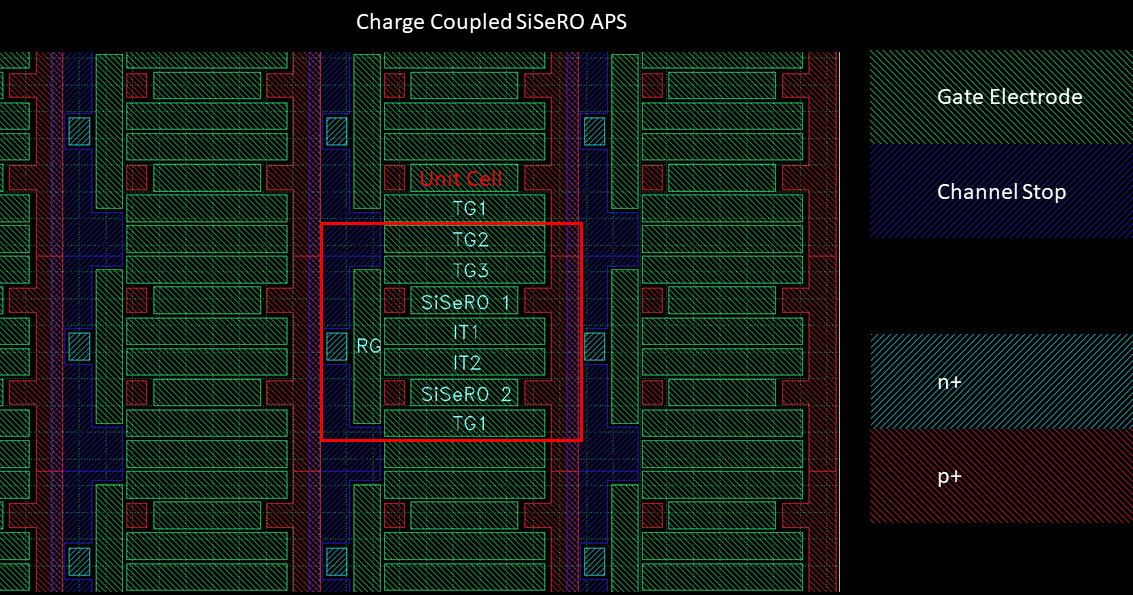}}
\end{subfigure}
\begin{subfigure}{0.425\textwidth}
    \includegraphics[width=\textwidth]{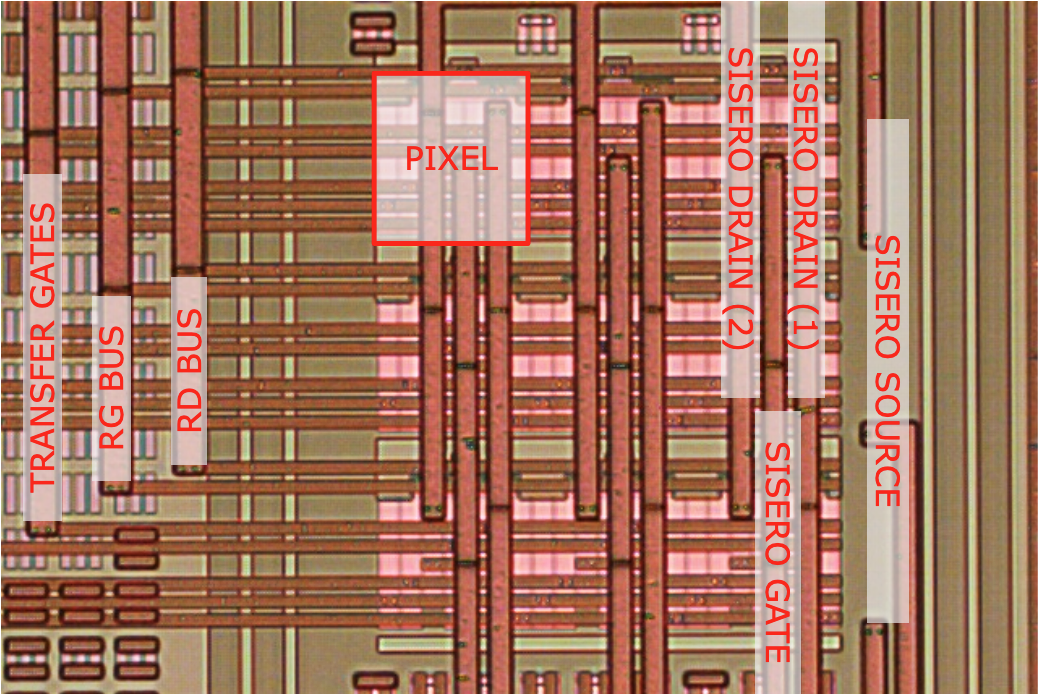}
    % \label{fig:sisero_diagram} 
\end{subfigure}
\caption{Left: Layout sketch for a SiSeRO active pixel matrix\cite{chattopadhyayDevelopmentTestingIntegrated2025}. Each pixel has two SiSeRO outputs with dedicated transfer gates between them to enable RNDR. The pixels are bounded by a channel stop (blue) and barrier gate structure (TG1).  Right: A photograph of the first $3\times3$ pixel SiSeRO APS prototype. }
\label{fig:APS}  

\end{figure}

We have also fabricated the first $3\times3$ pixel SiSeRO matrix, an active pixel sensor (APS) prototype. By combining RNDR capabilities with the extremely fast readout speed of the APS architecture, large format SiSeRO APS devices can enable high-resolution time-domain studies of rapidly varying sources \cite{arcodiaProspectsTimeDomainMultiMessenger2024}, reduce photon pileup when observing bright sources\cite{davisEventPileupChargecoupled2001}, and enhance background event rejection, enabling the observation of extremely faint sources \cite{cruiseNewAthenaMissionConcept2025, gaudiHabitableExoplanetObservatory2020}. In addition, the absence of macro charge transfer in these devices makes them more radiation hard, minimizing the impact of charge transfer inefficiency from radiation-induced detector defects \cite{hornbeckTrappingMinorityCarriers1955, daiOverviewFormationRadiationInduced2025}. Figure \ref{fig:APS} shows a layout and picture of the APS prototype. Development of the readout electronics and testbed for these devices is currently in progress and will be discussed in future works. 

\section{readout electronics and testbed}
\label{sec:RO}

Both CCID93++ SiSeRO flavors share the same form factor, allowing us to create a streamlined test bed that can accomodate both configurations. 

\begin{figure} []
\centering
\begin{subfigure}{0.65\textwidth}
    %\raisebox{10mm}
    {\includegraphics[width=\textwidth]{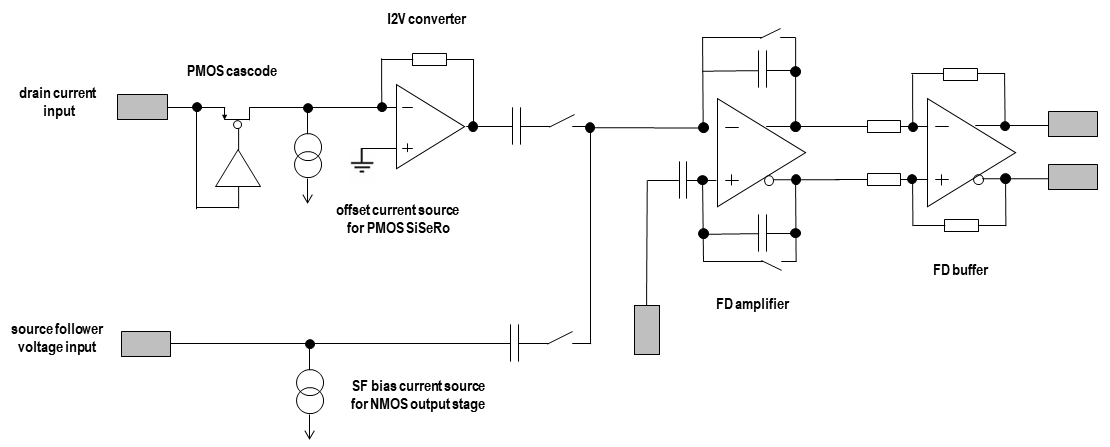}}
\end{subfigure}

\caption{The schematic for a single analog channel of the Multi-Channel Readout Chip (MCRC)\cite{herrmannMCRCV1Development2020}. The SiSeRO devices are read out in the drain current mode of the chip.}
\label{fig:MCRC}  

\end{figure}

Readout of both flavors is enabled by the Multi-Channel Readout Chip, an 8-channel application specific integrated circuit (ASIC) designed to provide high-speed readout for a fraction of the power consumption and footprint of traditional discrete readout electronics. In addition, the greater bandwidth of the MCRC allows us to operate SiSeROs at higher speeds \cite{chattopadhyayDevelopmentTestingIntegrated2025}.  As a proof of concept, we have successfully integrated a single-channel CCID93 detector with an MCRC and demonstrated noise and spectral performance comparable to that of the discrete readout electronics\cite{chattopadhyayDevelopmentTestingIntegrated2025}. To provide readout of all 16 channels of the CCID93++ detectors, two parallel MCRC chips are mounted on a custom daughter card. In drain current readout, the MCRC uses a current-to-voltage (I2V) converter to translate the single-ended SiSeRO transistor signal current into a voltage, which is then amplified and converted into a differential output signal ready for digitization. %The MCRC can also be configured in source follower mode to operate traditional pJFET-based CCD outputs.
Fig \ref{fig:MCRC} shows a schematic of a single analog channel in the MCRC. 

\begin{figure} []
\centering
\begin{subfigure}{0.48\textwidth}
    {\includegraphics[width=\textwidth]{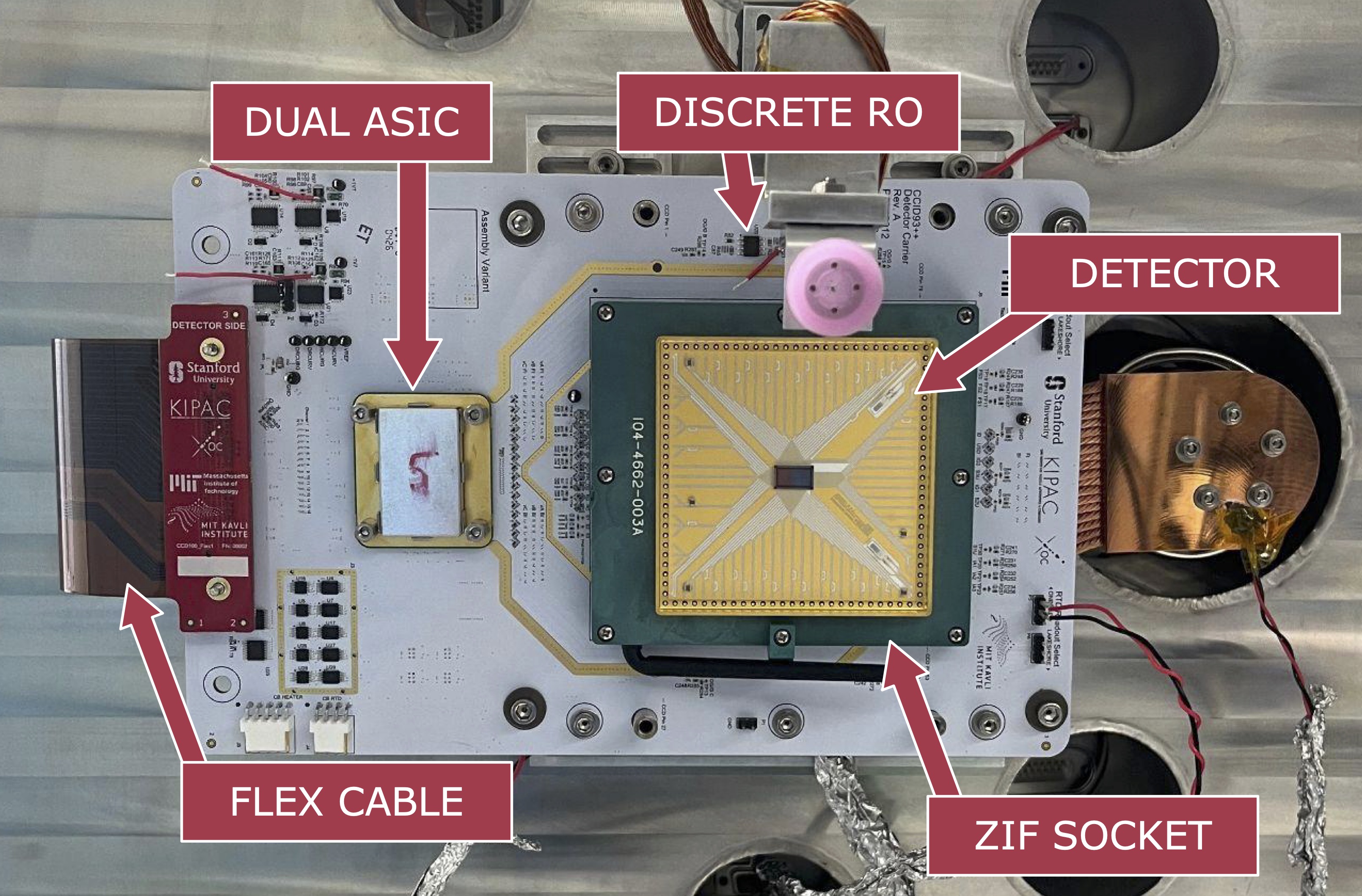}}
\end{subfigure}
\begin{subfigure}{0.423\textwidth}
    {\includegraphics[width=\textwidth]{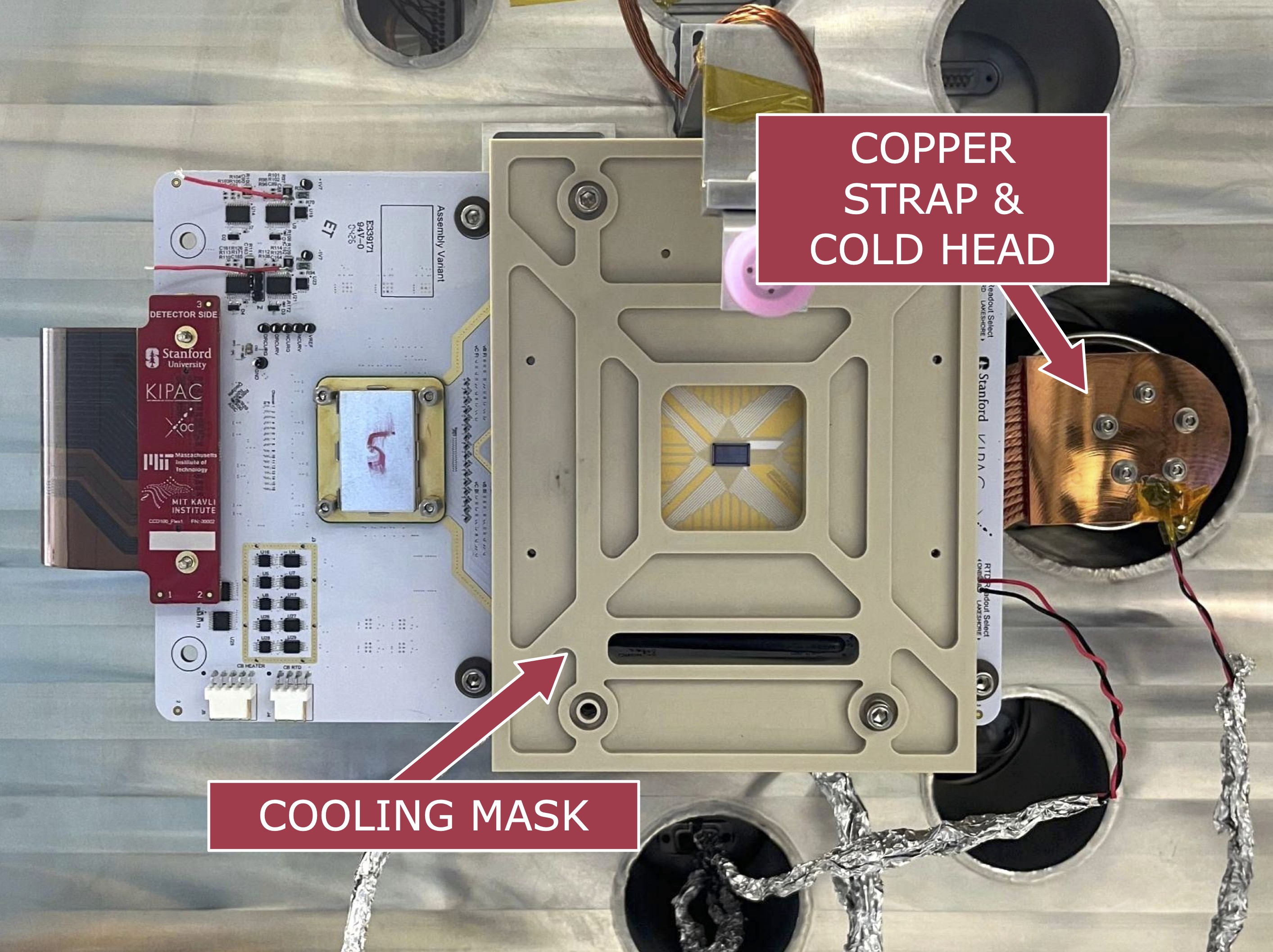}}
\end{subfigure}
\caption{Left: The CCID93++B readout assembly mounted to the vacuum chamber door of the testbed. The 16 detector outputs are read out by two Multi-Channel Readout Chip (MCRC) ASICs. Voltages and clock signals from the Archon Controller back-end electronics module are supplied to the detector through a custom interconnect board and flex cable. Right: The full CCID93++ mechanical assembly. The detector package is spring clamped to an aluminum block and copper thermal strap, which provide thermal contact to an Edward cryo-cooler coldhead for cooling to $<168\mathrm{K}$ }
\label{fig:teststand_inside}  

\end{figure}

The dual-MCRC daughter card is installed on a custom pre-amplifier board designed to accommodate all CCID93++ flavors (Figure \ref{fig:teststand_inside}). In addition to the MCRC, the pre-amplifier hosts one readout channel of discrete electronics for debugging and performance characterization. Five high-speed onboard clock drivers are used to clock the summing well (SW), output gate (OG), SiSeRO gate (S), and reset gate (RG) in each SiSeRO, along with an injection diode (ID) that can provide test signals to the detector imaging area. The MCRC DR input regulates the SiSeRO transistor drain terminal voltage to a fixed value of 1V. To control the operating point, the pre-amplifier board has been designed such that the MCRC sits in its own isolated power domain, which can be voltage shifted in order to accommodate a span of SiSeRO operating points. 

Control of the readout is provided by an FPGA-based Archon controller from Semiconductor Technology Associates, Inc\cite{bredthauerArchonModernController2014}. The Archon controller serves as our back-end electronics module, providing voltages and clocks to the MCRC and the detector while simultaneously digitizing the raw analog waveforms and running the correlated-double-sampling (CDS) event extraction algorithm  to provide pixel images of the detector image \cite{whiteCharacterizationSurfaceChannel1974,barbeImagingDevicesUsing1975}. Voltages and clock timing can be programmed through a graphical user interface (GUI), which also provides real-time voltage and current consumption diagnostics. 

\begin{figure} [b]
\centering
\begin{subfigure}{0.75\textwidth}
    % \raisebox{10mm}
    {\includegraphics[width=\textwidth]{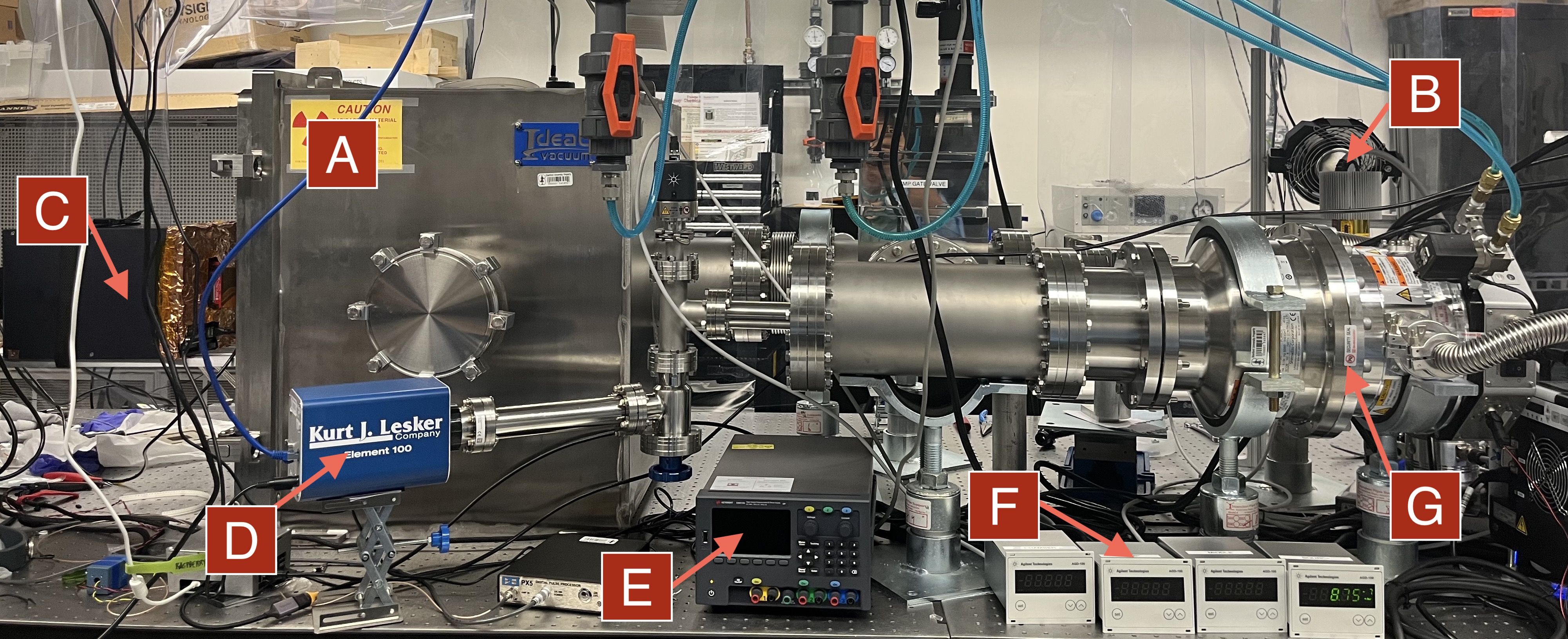}}
\end{subfigure}

\caption{The XOC Gen2.0 Beamline. The vacuum chamber (A) can be seen on the far left, and the XRF source module (B) can be seen on the far right. Additional electronics equipment include the Archon controller (C), a residual gas analyzer for vacuum health monitoring (D), the detector temperature monitor and PID system (E), vacuum gauge displays (F), and pump system (G).}
\label{fig:teststand}  

\end{figure}

During testing, the readout chain is installed in the XOC Gen2.0 vacuum beamline chamber (Figure \ref{fig:teststand})\cite{panDesignDevelopmentCommissioning2025}. The detector and readout board are mounted to the inside of a $20\times20\times24$ inch vacuum test chamber and are connected via a custom vacuum-potted flex cable and interconnect board to the Archon controller. The detector is cooled to $<168$K ($-105^\circ\mathrm{C}$) with an Edwards cryo-cooler to reduce thermal noise and dark current (Figure \ref{fig:teststand_inside}).

The beamline is equipped with a radioactive $^{55}\mathrm{Fe}$ source for simple detector noise and spectral characterization tests, and an X-ray fluorescence (XRF) module at the other end of the beamline. This XRF module provides bright, mono-energetic X-ray fluorescence lines between 0.5 - 9 keV. More details on the design of the beamline and test bed can be found in Pan et al. 2025\cite{panDesignDevelopmentCommissioning2025}. 

\section{CCID93++B Initial Testing}
\label{sec:testing}

\begin{figure} []
\centering
\begin{subfigure}{0.6\textwidth}
    {\includegraphics[width=\textwidth]{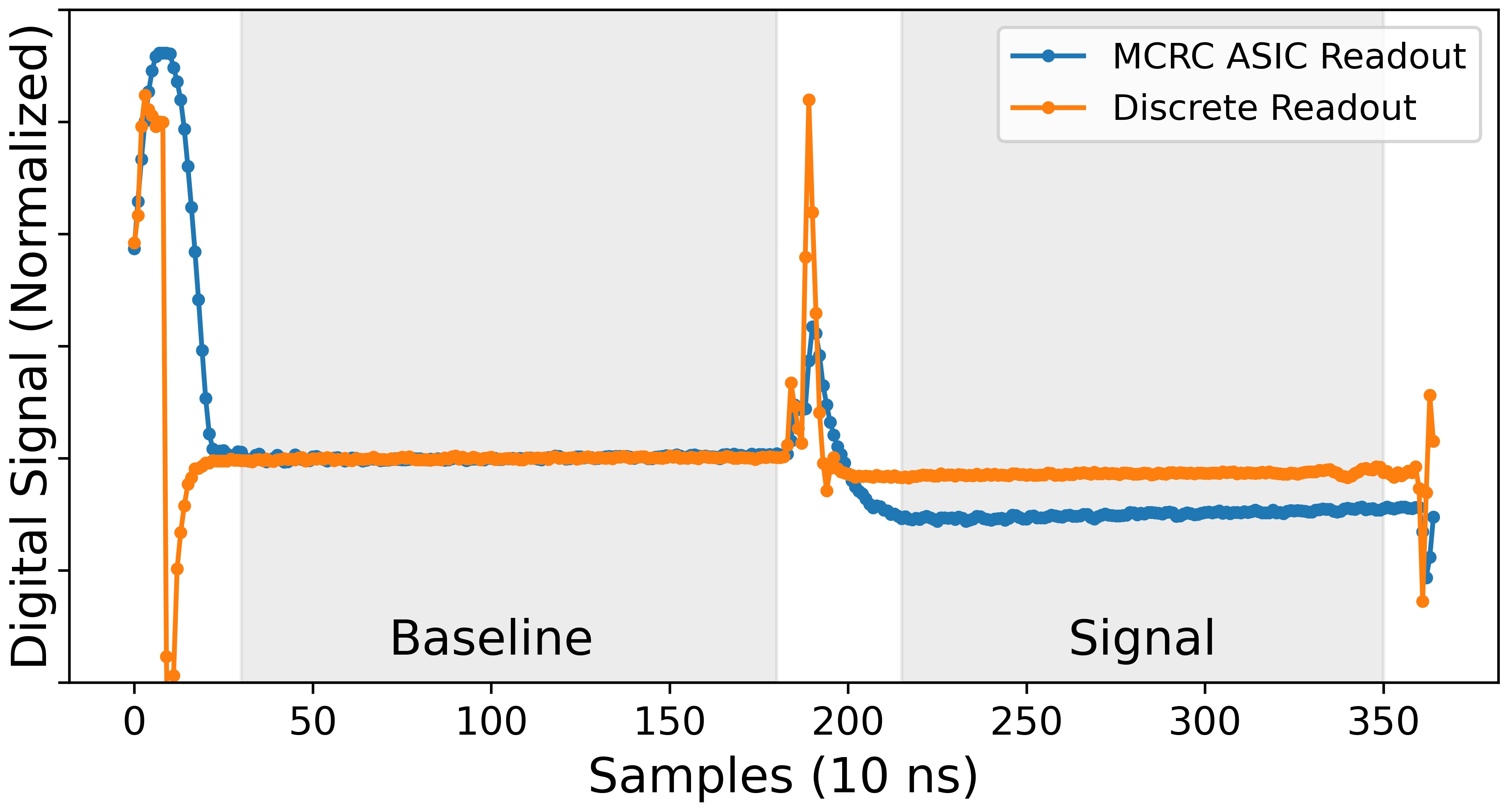}}
\end{subfigure}

\caption{Raw digital waveform of a single pixel from the discrete electronics and MCRC ASIC. The waveforms have been normalized for comparison purposes. During readout, a measurement is done with correlated double sampling: the pixel charge is the difference between the averaged baseline and signal value (gray shaded region) }
\label{fig:waveform}  

\end{figure}

Commissioning of the CCID93++B parallel SiSeRO is currently in progress. To provide a point of comparison, one detector channel is read out using discrete electronics, while the remaining 15 channels are operated by the MCRC. Figure \ref{fig:waveform} compares the raw digital waveform produced by the discrete electronics and MCRC. 

On the discrete electronics, we measure the transistor current of the SiSeRO p-MOSFET by manually probing the output of the I2V converter with a digital oscilloscope.
We also developed a automated process to measure all four groups of SiSeRO variants simultaneously. The transistor current of each group can be derived from the current draw of the SiSeRO source terminal bias voltage channel, as measured by the Archon controller. Because each group of outputs is biased by a shared voltage channel, the measured current draw is the sum of all four transistor currents plus the intrinsic current draw of the Archon controller itself. We assume that the outputs in a given variant group are roughly similar to approximate the transistor current of a single output. This process can be programmed to sweep across multiple bias points to quickly characterize the device. 

All 16 SiSeRO amplifiers were individually tested by observing the response of the I2V converter when changing the SiSeRO gate voltage bias. Fifteen out of sixteen amplifiers responded to the gate voltage. The cause behind the unresponsive amplifier is currently being investigated. 

\begin{figure} []
\centering
    {\includegraphics[width=0.9\textwidth]{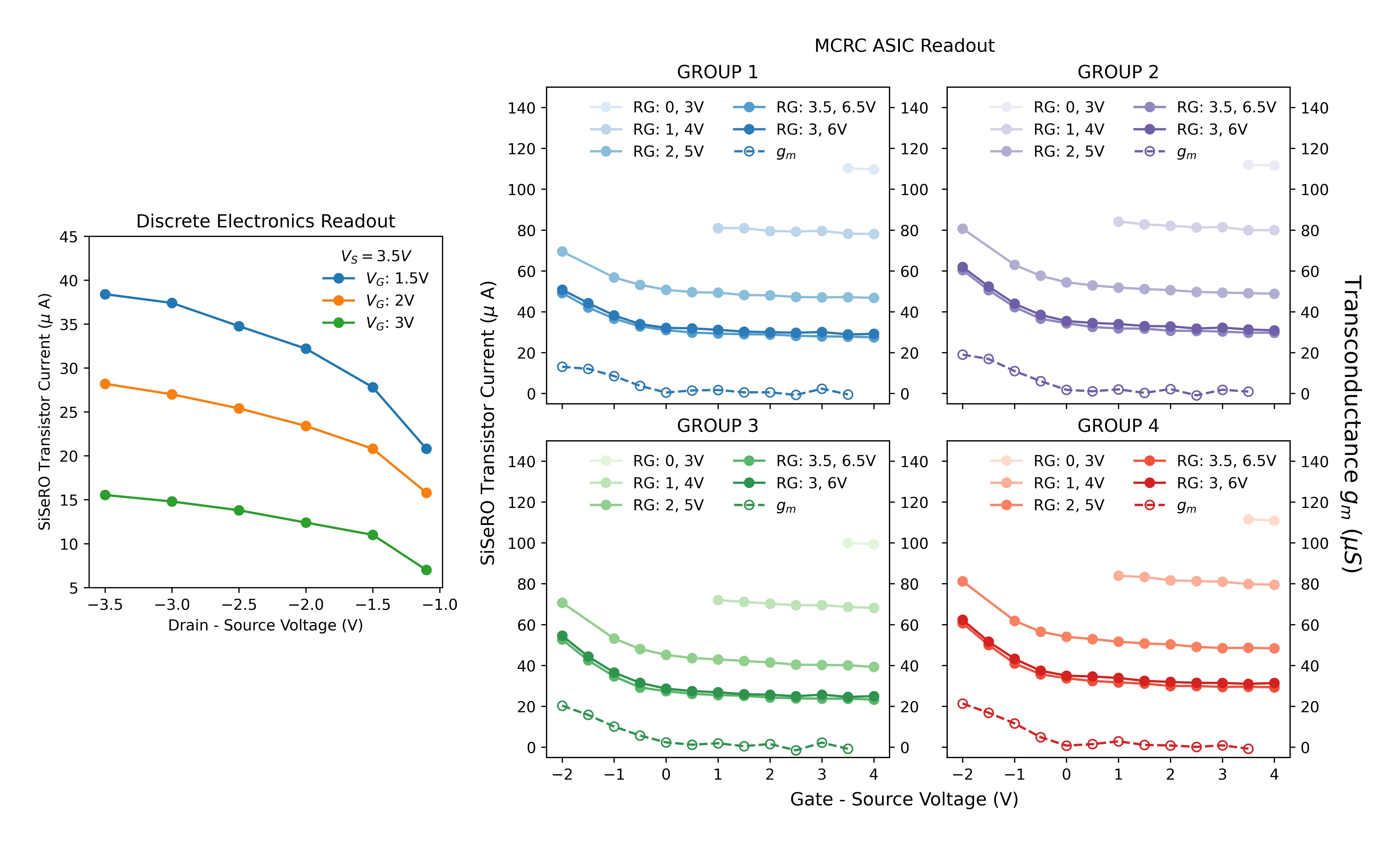}}

\caption{Left: Characteristic transistor curves (Drain-source voltage versus input current) for the discrete readout of one CCID93++B channel. The source voltage was kept at 3.5 V. Curves show the input current for gate voltages of 1.5 V (blue), 2 V (orange) and 3 V (green). Right: Transistor characterization curves (Gate-source voltage versus input current) for the MCRC ASIC readout. Each curve shows the input current for a single channel. For these tests, the source and drain voltage are kept at 4 V and 1 V respectively. A range of transistor gate and reset gate (RG) voltages were swept over. The dotted line shows the output transconductance at RG = 3, 6 V.}
\label{fig:I2V}  

\end{figure}

\begin{table}[b]

    \centering
    \begin{tabular}{|c|l|c|}
        \hline
         &\textbf{SiSeRO Variant} & \textbf{g$_m$ ($\mu$S)}\\
         \hline
         1 &Legacy & 13.08\\
         \hline
         2 &Shifted internal gate& 19.00\\
         \hline
         3 & Increased p-MOSFET isolation & 20.29\\
         \hline
         4 & Reduced p-MOSFET gate resistance & 21.32\\
         \hline \hline
         & \textbf{Average} & 18.42\\
        \hline
    \end{tabular}
    \caption{SiSeRO output variant transconductance.}
    \label{tab:transconductance}
\end{table}
Figure \ref{fig:I2V} shows the characteristic input current versus source-drain voltage curve (left) of as measured from the discrete electronics (i.e. Channel 1) (left) and transistor transfer characteristic curves (right) as measured on the MCRCs (i.e. Channels 2-15). For the discrete electronics measurements, the transistor source voltage was kept at 3.5 V, and the output current was measured while sweeping the drain voltage from 0 - 2.5 V for a number of fixed gate voltages. The transistor behaves as expected: hole current increases with the increase in the source-drainvoltage difference. 
For the MCRC ASIC, the transistor source and drain are biased at 4 V and 1 V, respectively, while sweeping the gate voltage from 2 to 8 V with our automated program. Missing data points indicate where the Archon was not able to apply the bias values, likely because the configuration exceeded the current draw limits of the system. Again, we observe the expected dependence of the transistor current on the gate voltage. We also observe that the baseline current heavily depends on the voltages of the clocked reset gate signal. As we have observed in previous SiSeRO detectors, improper biasing can lead to leakage paths around the transistor, increasing the overall current in the output  \cite{chattopadhyayFirstResultsSiSeRO2022}. The high current we are observing in this device is likely due to a large offset current caused by parasitic leakage between the output gate (OG) and reset gate (RG). 

Table \ref{tab:transconductance} shows the average transconductance for each SiSeRO output group at a nominal bias point of SiSeRO source: 4 V; drain: 1 V; gate: 2 V. We measure a maximum transconductance of 21.32 $\mu$S in group 4, the variant with reduced resistance in the p-MOSFET gate. On average, the transconductance is 18.42 $\mu$S.

We are currently performing optimization tests to find the best bias voltages to operate the SiSeRO outputs. We are testing different bias voltages for the output gate (OG) and the clocked reset gate (RG High \& Low) to minimize the offset current of the output.

\section{Summary and Future Plans}
\label{sec: Summary}

Single electron Sensitive Read Out (SiSeRO) is a novel readout technology targeting the technology gap of fast, low noise, large-format detectors for future flagship observatories in the X-ray to optical wavelength bands. First generation SiSeROs have demonstrated sub-electron read noise capabilities and show no evidence of any excess noise out to high numbers of RNDR cycles. 

We are currently testing the second generation of SiSeRO detectors to continue maturing the SiSeRO technology. These devices come in two variants: (1) a detector with 16 parallel SiSeROs, offering a 16$\times$ increase in readout speed. These SiSeRO outputs are fabricated with variations in the internal gate geometry, level of isolation between the MOSFET and surrounding structures, and MOSFET gate resistance, allowing us to efficiently optimize the SiSeRO design for gain, noise, and readout speed. (2) a MAS SiSeRO device with 16 identical SiSeROs in series. This configuration can achieve a $4\times$ reduction in readout noise, and will be used to optimize the RNDR capabilities of the SiSeRO design.  

We have designed and commissioned an electronics readout chain for characterizing these devices. We utilize the MCRC ASIC to provides low-power and low-noise readout of the detectors. Commissioning of the parallel SiSeRO device (the CCID93++B) is currently in progress in the XOC Gen2.0 X-ray vacuum beamline, which contains a new testbed to house the readout electronics and cool the detector. All but one of the 16 SiSeRO amplifiers are currently working. The maximum transconductance across all amplifiers is 21.32 $\mu$S; the average transconductance is 18.42 $\mu$S.

To explore our long-term goal of developing a SiSeRO active pixel sensor (APS), we have fabricated the first $3\times3$ pixel matrix as a proof of concept. This prototype will lay the groundwork for the development of large-scale SiSeRO APS matrices for future flagship astrophysics programs. Commissioning and characterization of this device will begin later this year.

\acknowledgments 
This work has been supported by the NASA APRA grant 80NSSC22K1921 and 80NSSC25K7957.

\bibliography{references,additionalbibs}
% \bibliography{additionalbibs}
\bibliographystyle{spiebib} % makes bibtex use spiebib.bst

\end{document}